\documentclass[a4paper]{jpconf}
\usepackage{graphicx}
\begin{document}
\title{Double beta decay experiments: past and present achievements}

\author{Alexander Barabash}

\address{Institute of Theoretical and Experimental Physics, B.Cheremushkinskaya 25,
 117218 Moscow, Russia}

\ead{barabash@itep.ru}

\begin{abstract}
A brief history of double beta decay experiments is presented. The best currently running experiments
 (NEMO-3 and CUORICINO) and their
latest results are described. The best measurements and limits for the $2\nu\beta\beta$,
 $0\nu\beta\beta$ and $0\nu\chi^{0}\beta\beta$
are summarized. 
\end{abstract}

\section{Historical introduction}

\subsection{Theory}

The double beta decay problem arose practically immediately after the appearance 
of W. Pauli's neutrino hypothesis 
in 1930 and the development of $\beta$-decay theory by E. Fermi in 1933. 
In 1935 M. Goeppert-Mayer identified for the first time 
the possibility of two neutrino double beta decay, 
in which there is a transformation of an (A, Z) 
nucleus to an (A, Z+2) nucleus that is accompanied by the emission of two electrons 
and two anti-neutrinos \cite{GOE35}:

\begin{equation}
(A,Z) \rightarrow (A,Z+2) + 2e^{-} + 2\tilde \nu                     
\end{equation}

It was demonstrated theoretically by E. Majorana in 1937 \cite{MAJ37} that if one 
allows the existence of only one type of neutrino, which has no antiparticle 
(i.e. $\nu \equiv\tilde\nu$) , then the conclusions of $\beta$-decay theory 
are not changed. In this case one deals with a Majorana neutrino. 
In 1939 B. Farry introduced a scheme of neutrinoless double beta decay 
through the virtual state of intermediate nuclei \cite{FUR39}: 

\begin{equation}
(A,Z) \rightarrow (A,Z+2) + 2e^{-}                                 
\end{equation}

In 1981 a new type of neutrinoless decay with Majoron emission was introduced \cite{GEO81}:

\begin{equation}
(A,Z) \rightarrow (A,Z+2) + 2e^{-} + \chi^{0}                     
\end{equation}

\subsection{Experiment} 

The first experiment to search for $2\beta$-decay was done in 1948 using Geiger counters. 
In this experiment a half-life limit for $^{124}$Sn was established, 
$T_{1/2} > 3\cdot 10^{15}$
 y \cite{FIR48}. During the period 1948 to 1965 $\sim$ 20 experiments were carried out with a sensitivity 
to the half-life on the level $\sim 10^{16}-10^{19}$ y (see reviews \cite{LAZ66,HAX84}). The $2\beta$-decay 
was thought to have been "discovered" a few times, but each time it was not confirmed by new 
(more sensitive) measurements. The exception was the geochemical experiment 
in 1950 where two neutrino double beta decay of $^{130}$Te was really detected \cite{ING50}. 
     
At the end of the 1960s and beginning of 1970s significant progress in the 
sensitivity of double beta decay experiments was realized. E. Fiorini et al. 
carried out experiments with Ge(Li) detectors and established a limit on 
neutrinoless double beta decay of $^{76}$Ge, $T_{1/2} > 5\cdot 10^{21}$ y \cite{FIO73}. 
Experiments with $^{48}$Ca and $^{82}$Se using streamer chamber with a magnetic field and 
plastic scintillators were done by C. Wu's group and led to impressive limits of 
$2\cdot 10^{21}$ y \cite{BAR70} and $3.1\cdot 10^{21}$ y \cite{CLE75} respectively. During 
these years many sensitive geochemical experiments were done and 
$2\nu\beta\beta$ decay of $^{130}$Te, $^{128}$Te and $^{82}$Se were 
detected (see reviews \cite{KIR83,HAX84,MAN86}). 
      
The important achievements in the 1980s were connected with the first evidence of 
two neutrino double beta decay in direct counting experiments. This was 
done by M. Moe's group for $^{82}$Se using a TPC \cite{ELL87}. There was also the first use 
of semiconductor detectors made of enriched Ge in the ITEP-ErPI experiment \cite{VAS90}.
     
During the 1990s the two neutrino decay process was detected in many experiments 
for different nuclei (see \cite{BAR02,BAR04}), two neutrino decay to an excited state 
of the daughter nuclei was also detected \cite{BAR95}. In addition, the sensitivity to $0\nu\beta\beta$ 
decay in experiments with $^{76}$Ge (Hidelberg-Moscow \cite{KLA01} and IGEX \cite{AAL02}) 
was increased up to $\sim 10^{25}$ y.  
     
Since 2002 the progress in double beta decay searches has been 
connected with the two best present experiments, NEMO-3 and CUORICINO (see Section 2).

\section{Best present experiments}

\subsection{NEMO-3}

This is a tracking experiment that, in contrast to experiments
with $^{76}$Ge, detects not only the total energy deposition,
but also other parameters of the process, including the energy of
the individual electrons, angle between them,
and the coordinates of the event in the source plane.  
Since June of 2002, the NEMO-3 detector \cite{ARN05a}
has operated at the Frejus Underground Laboratory (France)
located at a depth of 4800 m w.e. The detector has a cylindrical
structure and consists of 20 identical sectors.
A thin (about 30-60 mg/cm$^{2}$) source containing double beta decaying
nuclei and having a total area of 20 m$^{2}$ and a weight
of up
to 10 kg was placed in the detector.  
The energy of the electrons is measured by plastic
scintillators (1940 individual counters), while the tracks
are reconstructed on the basis of information obtained in the
planes of Geiger cells (6180 cells) surrounding the source
on both sides.  In addition, there is a magnetic field of 25 G
parallel to the detector axis which is created by a solenoid
surrounding the detector.

\begin{figure*}
\begin{center}
\includegraphics[scale=0.3]{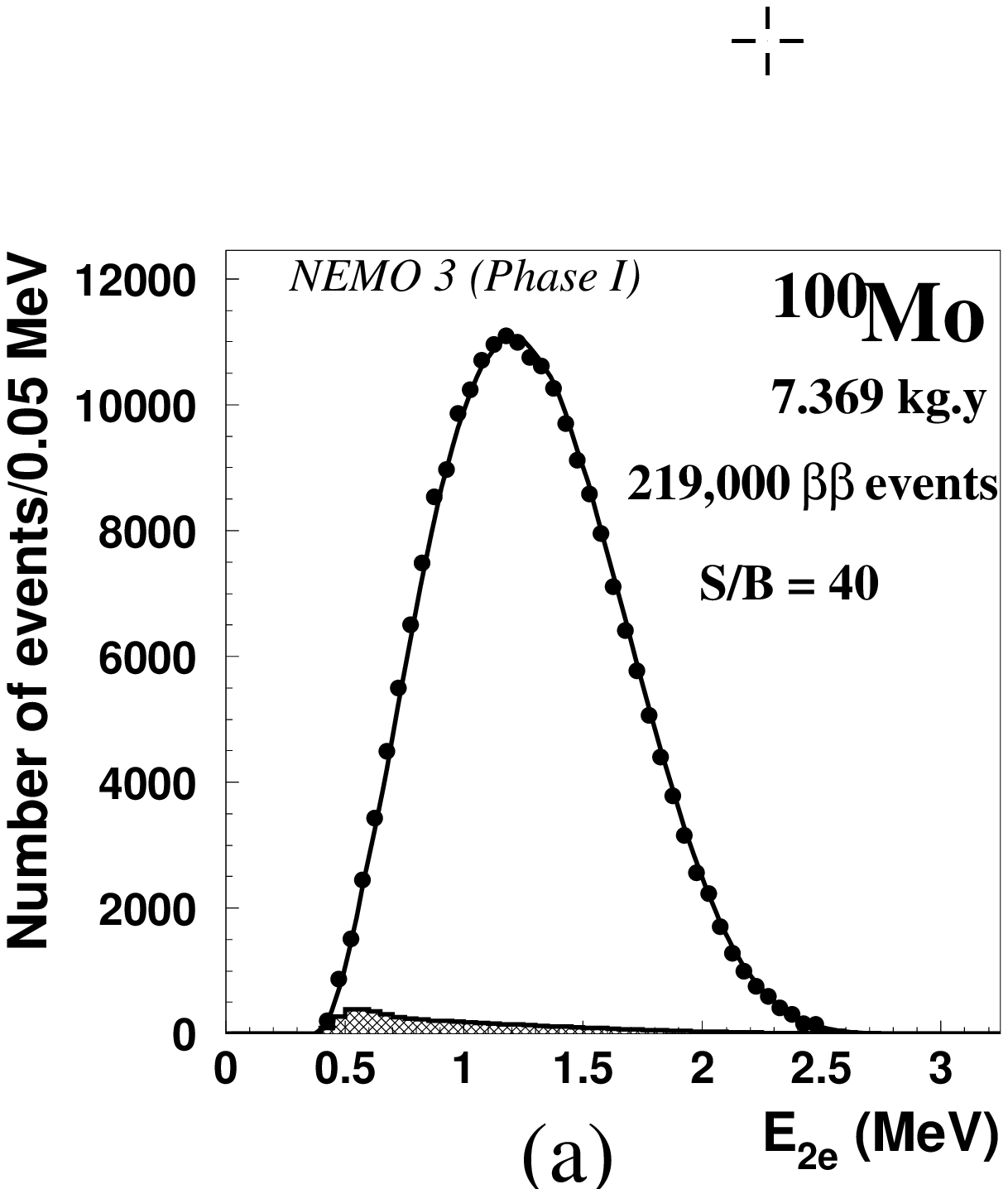}
\includegraphics[scale=0.3]{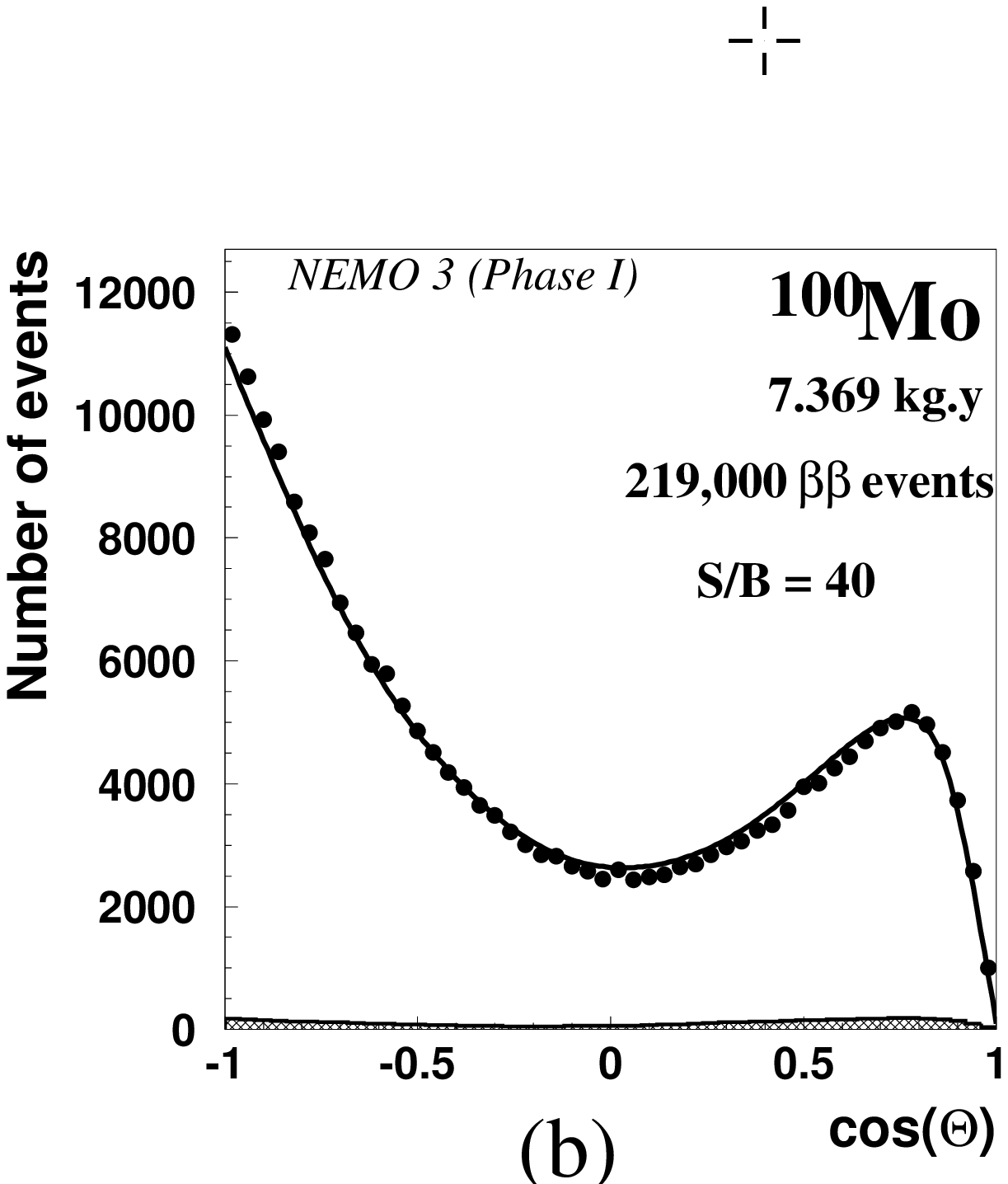}
\includegraphics[scale=0.3]{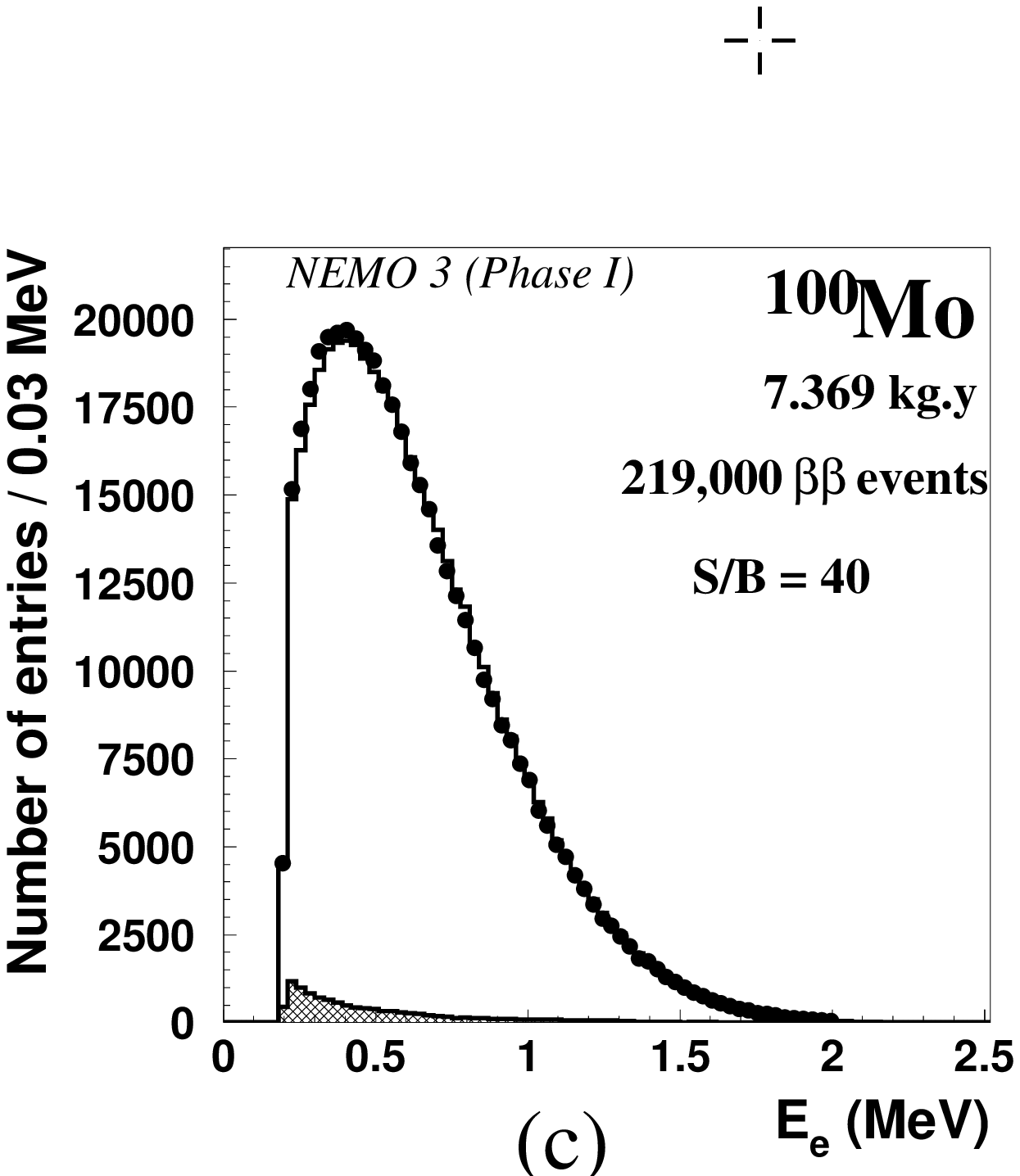}
\caption{\label{fig:figure1} (a) Energy sum spectrum of the two
electrons, (b) angular
distribution of the two electrons and (c) single energy spectrum
of the electrons, after background
subtraction from $^{100}$Mo with 7.369 kg$\cdot$years  
exposure \cite{ARN05}.}
\end{center}
\end{figure*}

At the
present time, studies are being performed
on seven double beta decay isotopes; these are $^{100}$Mo (6.9 kg), $^{82}$Se (0.93 kg), 
$^{116}$Cd (0.4 kg),
$^{150}$Nd (37 g), $^{96}$Zr (9.4 g), $^{130}$Te (0.45 kg), 
and $^{48}$Ca (7 g). 

Fig.~\ref{fig:figure1} displays the
spectrum of $2\beta(2\nu)$ events
in $^{100}$Mo that were collected over 389 days
\cite{ARN05}. 
The total number of events is about 219,000, which is much greater
than the total statistics of all of the preceding
experiments. Given the calculated values for the detection
efficiencies for
$2\beta(2\nu)$ events, the following half-life value was
obtained for $^{100}$Mo:

\begin{equation}
T_{1/2}(^{100}Mo;2\nu) = [7.11 \pm 0.02(stat) \pm 0.54(syst)
]\cdot 10^{18}\; y\
\end{equation}

Using 690 days of data the following
limits on neutrinoless double beta decay of $^{100}$Mo and
$^{82}$Se (mass mechanism; 90\% C.L.) have been obtained:
$> 5.8\cdot 10^{23}$ y and $> 2.1\cdot 10^{23}$ y.

The corresponding limits on $\langle m_{\nu} \rangle$ 
are presented in Table 2.

\subsection{CUORICINO}

This program is the first stage of the larger experiment CUORE. The experiment
is running at the Gran Sasso Underground Laboratory. The detector
consists of low-temperature devices
based on $^{nat}$TeO$_{2}$  crystals. The detector consists of 62
individual crystals, their total mass being 40.7 kg.
The energy resolution is $\sim$ 8 keV (FWHM) at 2.6 MeV.

\begin{figure}
\begin{center}
\includegraphics[scale=0.4]{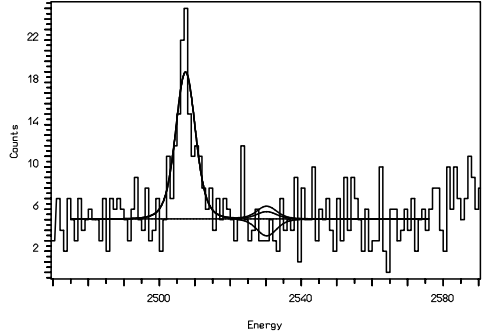}
\caption{\label{fig:fig2} The sum spectra of all $^{nat}$TeO$_{2}$ crystals in the
region
of the $2\beta(0\nu)$ energy of $^{130}$Te \cite{CAP06}.}
\end{center}
\end{figure}

The experiment has been running
 since March of 2003. The summed spectra of all crystals in the
region
of the $2\beta(0\nu)$ energy is shown in Fig.~\ref{fig:fig2}.
The total exposure is 8.3 $kg \cdot$ y  ($^{130}$Te). The
background
at the energy of the $2\beta(0\nu)$ decay is 0.18 keV$^{-1}\cdot
kg^{-1} \cdot y^{-1}$.  No peak is evident and the limit is
$T_{1/2} > 2.4\cdot 10^{24}$ y (90\% CL) \cite{CAP06}. 
The corresponding limits on $\langle m_{\nu} \rangle$ 
are presented in Table 2.

\section{Best achievements}

\subsection{Two neutrino double beta decay}

As discussed above this decay was first recorded in 1950 in a geochemical experiment
with $^{130}$Te \cite{ING50}. In 1967,
it was also found for $^{82}$Se \cite{KIR67}.
Attempts to observe this decay in a direct measurement employing
counters were unsuccessful for
a long time. Only in 1987 could M. Moe, who used a time-projection
chamber (TPC), observe $2\beta(2\nu)$ decay
in $^{82}$Se
for the first time \cite{ELL87}. Within the next few years,
experiments employing counters were able to detect
$2\beta(2\nu)$
decay in many nuclei. In $^{100}$Mo \cite{BAR95,BAR99,BRA01}, and
$^{150}$Nd \cite{BAR04a}
$2\beta(2\nu)$ decay to the $0^{+}$ excited state of the
daughter nucleus was also recorded. The
$2\beta(2\nu)$ decay of $^{238}$U was detected in a radiochemical
experiment \cite{TUR91}, and in a geochemical experiment for the
first time the ECEC process was detected in $^{130}$Ba
\cite{MES01}. Table 1 displays the present-day averaged and
recommended values of
$T_{1/2}$(2$\nu$) from \cite{BAR06}.

\begin{table}[ht]
\label{Table1}
\caption{Average and recommended $T_{1/2}(2\nu)$ values (from
\cite{BAR06}).}
\vspace{0.5cm}
\begin{center}
\begin{tabular}{cc}
\br
Isotope & $T_{1/2}(2\nu)$ \\
\hline
$^{48}$Ca & $4.2^{+2.1}_{-1.0}\cdot10^{19}$ \\
$^{76}$Ge & $(1.5 \pm 0.1)\cdot10^{21}$ \\
$^{82}$Se & $(0.92 \pm 0.07)\cdot10^{20}$ \\
$^{96}$Zr & $(2.0 \pm 0.3)\cdot10^{19}$ \\
$^{100}$Mo & $(7.1 \pm 0.4)\cdot10^{18}$ \\
$^{100}$Mo-$^{100}$Ru$(0^{+}_{1})$ & $(6.8 \pm 1.2)\cdot10^{20}$ \\
$^{116}$Cd & $(3.0 \pm 0.2)\cdot10^{19}$\\
$^{128}$Te & $(2.5 \pm 0.3)\cdot10^{24}$ \\
$^{130}$Te & $(0.9 \pm 0.1)\cdot10^{21}$ \\
$^{150}$Nd & $(7.8 \pm 0.7)\cdot10^{18}$ \\
$^{150}$Nd-$^{150}$Sm$(0^{+}_{1})$ & $1.4^{+0.5}_{-0.4}\cdot10^{20}$
\\
$^{238}$U & $(2.0 \pm 0.6)\cdot10^{21}$  \\
$^{130}$Ba; ECEC(2$\nu$) & $(2.2 \pm 0.5)\cdot10^{21}$  \\
\br
\end{tabular}
\end{center}
\end{table}

\subsection{Neutrinoless double beta decay}

In contrast to two-neutrino decay, neutrinoless double-beta decay
has not yet been observed
\footnote{A fraction of the  Heidelberg-Moscow
Collaboration published a "positive" result for $^{76}$Ge, 
$T_{1/2} \approx 1.2\cdot 10^{25}$ y \cite{KLA04} (see
table 2). The Moscow portion of the
Collaboration does not agree with this conclusion \cite{BAK03} and
there is independent criticism
of this result (see, for example \cite{STR05}). Thus at the present time this
"positive" result is not accepted by the
"2$\beta$ decay community" and it has to be checked by new
experiments.}. 

The present-day constraints on the existence of
$2\beta(0\nu)$ decay are presented in table 2. In calculating constraints
on $\langle m_{\nu} \rangle$, the
nuclear matrix elements from \cite{SIM99,STO01,CIV03} were used (3-d column).  
In column four, limits on $\langle m_{\nu} \rangle$ are given,
which were obtained using the NMEs (QRPA and RQRPA calculations)
from a recent paper \cite{ROD05}. In this paper $g_{pp}$ values ($g_{pp}$ is 
a parameter in QRPA theory) were fixed using experimental 
half-life values for $2\nu$ decay and then
NME(0$\nu$) were calculated. 
The authors \cite{ROD05} analyzed the results of all existing QRPA calculations which 
demonstrates that their approach
gives the most accurate and reliable values for NMEs. There is criticism to this claim 
in \cite{SUH05}). 

\begin{table}[ht]
\label{Table2}
\caption{Best present results on $2\beta(0\nu)$ decay (limits at
90\% C.L.). $^{*)}$ See text.}
\vspace{0.5cm}
\begin{center}
\begin{tabular}{ccccc}
\br
Isotope & $T_{1/2}$, y & $\langle m_{\nu} \rangle$, eV & $\langle
m_{\nu} \rangle$, eV & Experiment \\
& & \cite{SIM99,STO01,CIV03} & \cite{ROD05}  \\
\hline
$^{76}$Ge & $>1.9\cdot10^{25}$ & $<0.33-0.84$ & $<0.46-0.59$ & HM
\cite{KLA01} \\
& $\simeq 1.2\cdot10^{25}$(?)$^{*)}$ & $\simeq 0.5-1.3(?)$$^{*)}$ & $\simeq
0.6-0.7(?)$$^{*)}$ & Part of HM \cite{KLA04} \\
& $>1.6\cdot10^{25}$ & $<0.36-0.92$ & $<0.51-0.64$ & IGEX
\cite{AAL02} \\
\hline
$^{130}$Te & $>2.4\cdot10^{24}$ & $<0.4-0.8$ & $<0.7-1.3$ &
CUORICINO \cite{CAP06} \\
$^{100}$Mo & $>5.8\cdot10^{23}$ & $<0.6-0.9$ & $<2.0-2.7$ & NEMO-
3 (this work) \\
$^{136}$Xe & $>4.5\cdot10^{23}$ & $<0.8-4.7$ & $<2.8-5.6$ & DAMA
\cite{BER02} \\
$^{82}$Se & $>2.1\cdot10^{23}$ & $<1.2-2.5$ & $<2.3-3.2$ & NEMO-3
(this work) \\
$^{116}$Cd & $>1.7\cdot10^{23}$ & $<1.4-2.5$ & $<2.9-4.3$ &
SOLOTVINO \cite{DAN03} \\
\br
\end{tabular}
\end{center}
\end{table}

\subsection{Neutrinoless double beta decay with Majoron emission}

Table 3 displays the best present-day constraints for an "ordinary"
Majoron (spectral index n = 1). 
For "nonstandard" models of the Majoron (n = 2, 3 and 7) 
the strongest limits were obtained with the NEMO-3 detector \cite{ARN06}. 

\begin{table}[ht]
\label{Table3}
\caption{Best present limits on $2\beta(0\nu\chi^{0})$ decay
(ordinary Majoron) at 90\% C.L.}
\vspace{0.5cm}
\begin{center}
\begin{tabular}{cccc}
\br
Isotope & $T_{1/2}$, y & $\langle g_{ee} \rangle$, \cite{SIM99,STO01,CIV03}
 & $\langle
g_{ee} \rangle$, \cite{ROD05} \\  \\
\hline
$^{76}$Ge & $>6.4\cdot10^{22}$ \cite{KLA01} & $<(1.2-3.0)\cdot10^{-
4}$ & $<(1.9-2.3)\cdot10^{-4}$ \\
$^{82}$Se & $>1.5\cdot10^{22}$ \cite{ARN06} & $<(0.66-
1.4)\cdot10^{-4}$ & $<(1.3-1.8)\cdot10^{-4}$ \\
$^{100}$Mo & $>2.7\cdot10^{22}$ \cite{ARN06} & $<(0.4-
0.7)\cdot10^{-4}$ & $<(1.3-1.8)\cdot10^{-4}$ \\
$^{116}$Cd & $>8\cdot10^{21}$ \cite{DAN03} & $<(1.0-2.0)\cdot10^{-
4}$ & $<(2.3-3.5)\cdot10^{-4}$ \\
$^{128}$Te & $>2\cdot10^{24}$(geochem)\cite{MAN91} & $<(0.7-
1.6)\cdot10^{-4}$ & $<(1.7-2.8)\cdot10^{-4}$ \\
\br
\end{tabular}
\end{center}
\end{table}

\subsection{Improvements in experimental methods}

The dominate experimental achievements are connected with the following methods:

1)  Low background HPGe detectors made of enriched Ge (HM \cite{KLA01} and IGEX \cite{AAL02}).

2) Low background low temperature detectors (TeO$_2$ crystals; CUORICHINO \cite{ARN05}). 

3) Low background crystal scintillators ($^{116}$CdWO$_4$; SOLOTVINO \cite{DAN03}).

4) Low background tracking detectors (TPC \cite{LUE98}, NEMO-3 \cite{ARN05}). 

The key point is the level of the background, because high sensitivity experiments 
can only operate under low background conditions. The important achievements 
were done during the period of $2\beta$-decay searches. In table 4 one 
can find the best background levels reached which were obtained in the different experiments.
In 2-nd column, is the background $B$ in $(keV\cdot kg\cdot y)^{-1}$. 
For a better comparison of the experiments with different 
energy resolutions and efficiencies, a so called effective background value 
$\langle B \rangle$ has been introduced.

\begin{table}[h]
\caption{Lowest levels of background in double beta decay experiments.
$\Delta E$ is energy resolution (FWHM) in keV and $\eta$ is efficiency; $PSD$ 
is pulse shape discrimination.}
\begin{center}
\begin{tabular}{lll}
\br
Experiment & $B$, $(keV\cdot kg\cdot y)^{-1}$ & $\langle B \rangle$  = $B\cdot\Delta E/ \eta$,
 $(kg\cdot y)^{-1}$\\
\mr
HM \cite{KLA01}, IGEX \cite{AAL02}; $^{76}$Ge & $\sim 0.2$ & $\sim 0.8$\\
& $\sim 0.06 (PSD)$ & $\sim 0.25 (PSD)$\\
CUORICINO \cite{ARNA05}; TeO$_2$ & $\sim 0.18$ & $\sim 1.4$\\
NEMO-3 \cite{ARN05}; $^{100}$Mo & $\sim 0.001$ & $\sim 2.5$ \\
SOLOTVINO \cite{DAN03}; $^{116}$CdWO$_4$ & $\sim 0.037$ & $\sim 10$\\
TPC \cite{LUE98}; $^{136}$Xe & $\sim 0.02$ & $\sim 15$ \\
DAMA \cite{BER02}; $^{136}$Xe & $\sim 0.06$ & $\sim 30$ \\ 
\br
\end{tabular}
\end{center}
\end{table}

\section{Conclusion}

The principle achievements of past and present experiments are the following:

1) A conservative limit on the effective Majorana neutrino mass has been established as $< 0.9$ eV (90\% C.L.).

2) A conservative limit on the coupling constant of Majoron to neutrino (ordinary Majoron) 
has been established as  $<1.8\cdot 10^{-4}$ (90\% C.L.).

3) The two neutrino double beta decay has been detected in 10 nuclei. In addition this type of 
decay to the $0^+$ excited state of daughter nuclei has been detected (for $^{100}$Mo and 
$^{150}$Nd). Finally, the ECEC$(2\nu)$ process has been detected in geochemical experiment.


\section{References}

\begin{thebibliography}{<25>}

\bibitem{GOE35} 
Goeppert-Mayer M 1935 {\it Phys. Rev.} {\bf 48} 512
\bibitem{MAJ37}
Majorana E 1937 {\it Nuovo Cimento} {\bf 14} 171
\bibitem{FUR39}
Farry W H 1939 {\it Phys. Rev.} {\bf 56} 1184
\bibitem{GEO81}
Georgi H, Glashow S and Nussinov S 1981 {\it Nucl. Phys.} B {\bf 193} 297
\bibitem{FIR48}
Firemann E L 1948 {\it Phys. Rev.} {\bf 74} 1238
\bibitem{LAZ66}
Lazarenko V R 1966 {\it Uspechi} {\bf 90} 601
\bibitem{HAX84}
Haxton W C and Stephenson G J 1984 {\it Prog. Part. Nucl. Phys.} {\bf 12} 409
\bibitem{ING50}
Inghram M G and Reynold J H 1950 {\it Phys. Rev.} {\bf 78} 822
\bibitem{FIO73}
Fiorini E et al. 1973 {\it Nuovo Cimento} {\bf 13A} 747
\bibitem{BAR70}
Bardin R K Gollon P J Ullman J D and Wu C S 1970 {\it Nucl. Phys.} A {\bf 158} 337
\bibitem{CLE75}
Cleveland B T et al. 1975 {Phys. Rev. Lett.} {\bf 35} 757
\bibitem{KIR83}
Kirsten T 1983 {\it AIP Conference Proceedings} {\bf 96} 396
\bibitem{MAN86}
Manuel O K 1986 {\it Proc. Int. Work. Nuclear beta decays and the neutrino (Osaka)} 
(Singapure: World Scientific) p 71
\bibitem{ELL87}
Elliott S R Hahn A A and Moe M K 1987 {\it Phys. Rev. Lett.} {\bf 59} 2020
\bibitem{VAS90}
Vasenko A A et al. 1990 {\it Mod. Phys. Lett.} A {\bf 5} 1299
\bibitem{BAR02}
Barabash A S 2002 {\it Czech. J. Phys.} {\bf 52} 567 ({\it Preprint} nucl-ex/0203001)
\bibitem{BAR04}
Barabash A S 2004 {\it Phys. At. Nucl.} {\bf 67} 438
\bibitem{BAR95}
Barabash A S et al. 1995 {\it Phys. Lett.} B {\bf 345} 408 
\bibitem{KLA01}
Klapdor-Kleingrothaus H V et al. 2001 {\it Eur. Phys. J.} A {\bf 12} 147
\bibitem{AAL02}
Aalseth C E 2002 {\it Phys. Rev.} C {\bf 65} 092007
\bibitem{ARN05a}
Arnold R et al. 2005 {\it Nucl. Instr. Meth.} A {\bf 536} 79 
\bibitem{ARN05}
Arnold R et al. 2005 {\it Phys. Rev. Lett.} {\bf 95} 182302
\bibitem{CAP06}
Capelli S 2006 {\it Poster presentation at NEUTRINO'06 Conference (Santa Fe, USA)}
\bibitem{KIR67}
Kirsten T Gentner W and Schaeffer O A 1967 {\it Z. Phys.} {\bf 202} 273
\bibitem{BAR99}
Barabash A S et al. 1999 {\it Phy. At. Nucl.} {\bf 62} 2039
\bibitem{BRA01}
De Braeckeleer L et al. 2001  {\it Phys. Rev. Lett.} {\bf 86} 3510
\bibitem{BAR04a}
Barabash A S  2004 {\it Phys. At. Nucl.} {\bf 79} 10
\bibitem{TUR91}
Turkevich A L Economou T E and Cowan G A 1991 {\it Phys. Rev. Lett.} {\bf 67} 3211
\bibitem{MES01}
Meshik A P 2001 {\it Phys. Rev.} C {\bf 64} 035205
\bibitem{BAR06}
Barabash A S 2006 {\it Czech. J. Phys.} {\bf 56} 437 ({\it Preprint} nucl-ex/0602009) 
\bibitem{KLA04}
Klapdor-Kleingrothaus H V et al. 2004 {\it Phys. Lett.} B {\bf 586} 198
\bibitem{BAK03}
Bakalyarov A M et al. 2005 {\it Phys. Part. Nucl. Lett.} {\bf 2} 77 ({\it Preprint} 
hep-ex/03090160)
\bibitem{STR05}
Strumia A and Vissani F 2005 {\it Nucl. Phys.} B {\bf 726} 294
\bibitem{SIM99}
Simkovic F et al. 1999 {\it Phys. Rev.} C {\bf 60} 055502
\bibitem{STO01}
Stoica and Klapdor-Kleingrothaus H V 2001 {\it Nucl. Phys.} A {\bf 694} 269  
\bibitem{CIV03}
Civitarese O and Suhonen J 2003 {\it Nucl. Phys.} A {\bf 729} 867 
\bibitem{ROD05}
Rodin V A et al 2006 {\it Nucl. Phys.} A {\bf 766} 107
\bibitem{BER02}
Bernabei R et al. 2002 {\it Phys. Lett.} B {\bf 546} 23 
\bibitem{DAN03}
Danevich F A et al. 2003 {\it Phys. Rev.} C {\bf 67} 035501
\bibitem{SUH05}
Suhonen J 2005 {\it Nucl. Phys.} A {\bf 752} 53c 
\bibitem{ARN06}
Arnold R et al. 2006 {\it Nucl. Phys.} A {\bf 765} 483 
\bibitem{MAN91}
Manuel O K 1991 {\it J. Phys.} G {\bf 17} 221 
\bibitem{GUN97}
Gunther M et al. 1997 {\it Phys. Rev.} D {\bf 55} 54 
\bibitem{ARNA05}
Arnaboldi C et al. 2005 {\it Phys. Rev. Lett.} {\bf 95} 142501 
\bibitem{LUE98}
Luescher R et al. 1998 {\it Phys. Lett.} B {\bf 434} 407
 


\end{thebibliography}
\end{document}